\documentclass{article}

\usepackage{graphicx}

\newcommand{\Tr}{\mathrm{Tr}}
\newcommand{\const}{\mathrm{const}}

\newcommand{\cM}{\mathcal{M}}
\newcommand{\cN}{\mathcal{N}}
\newcommand{\cT}{\mathcal{T}}

\newcommand{\dd}{\mathrm{d}}

\begin{document}

\begin{center}
\vspace{12pt}
{ \large \bf The transfer matrix in four-dimensional Causal Dynamical Triangulations }\\[24pt]
{\sl Andrzej G\"{o}rlich}\\[12pt]
{\footnotesize
The Niels Bohr Institute, Copenhagen University\\
Blegdamsvej 17, DK-2100 Copenhagen \O , Denmark.\\
email: goerlich@nbi.dk
}
\vspace{24pt}
\end{center}

\begin{abstract}
\noindent Causal Dynamical Triangulations is a background independent approach to quantum gravity.
We show that there exists an effective transfer matrix
labeled by the scale factor which properly describes the evolution of the quantum universe.
In this framework no degrees of freedom are frozen, but,
the obtained effective action agrees with the minisuperspace model.
\end{abstract}


\section{Introduction}
In Causal Dynamical Triangulations (CDT) 
the quantum gravity path integral is regularized 
by approximating the spacetime geometry $g$ with a piecewise linear manifold $\cT$ \cite{Dyna, Reco}.
The building blocks of four dimensional CDT are four-simplices,
which form a simplicial manifold.
Imposed global proper-time foliation allows a well-defined Wick rotation.
Spatial slices, enumerated by a discrete time coordinate $t$,
are built of equilateral tetrehedra with link length $a_s$.
Time-like links have length $a_t$.
There are two kinds of simplices: 
$\{4, 1\}$ and $\{3, 2\}$ where the numbers denote the number of vertices in adjacent slices.
The regularized partition function $Z$ is written as a sum over causal triangulations $\cT$,
\[
  Z = \int \mathcal{D}[g] e^{i S^{EH}[g]} \,\rightarrow \, \sum_{\cT} e^{- S[\cT]}, \
S[\cT] = - K_0\ N_0 + K_4\ N_4  + \Delta\ (N_{41} - 6 N_0),
\]
where $K_0$, $K_4$ and $\Delta$ are bare coupling constants,
which are functions of $G, \lambda$ and $a_t, a_s$.
The discrete Regge action $S[\cT]$ is 
the Einstein-Hilbert action $S^{EH}[g] = \frac{1}{16 \pi G} \int \dd t \int \dd^3 x \sqrt{- g} (R - 2 \Lambda)$
evaluated on a simplicial manifold $\cT$ composed of 
$N_{0}$ vertices and
$N_{4}$ simplices, among them $N_{41}$ of type $\{4, 1\}$,

Using Monte Carlo simulations,
we can measure expectation values of observables within the CDT framework.
The simplest observable is the three-volume $n_t$
defined as the number of tetrahedra building slice $t$.
In the de Sitter phase
the average volume profile $\langle n_t \rangle \propto \cos^3(t / B)$
corresponds to a four-dimensional Euclidean de Sitter universe
which emerges dynamically as a background geometry.
Studies of the covariance matrix $\langle (n_t - \langle n_t \rangle) (n_{t'} -  \langle n_{t'}\rangle) \rangle$
showed that the fluctuations of $n_t$ 
are well described by the discretized minisuperspace action\cite{Plan},
\begin{equation}
	S[{n_t}] = \frac{1}{\Gamma} \sum_t \left( \frac{(n_{t+1} - n_t)^2}{n_{t+1} + n_t} + \mu n_t^{1/3} - \lambda  n_t  \right).
	\label{eq:SMiniD}
\end{equation}

\section{The transfer matrix}

The CDT model is completely determined by a transfer matrix $\cM$ ,
\begin{equation}
	Z = \sum_{\mathcal{T}} e^{- S^{[\mathcal{T}]}} = \Tr \cM^T, \quad \langle \tau_1 | \cM | \tau_2 \rangle = \sum_{\cT|_{\tau_1, \tau_2}} e^{-S[\cT] }.
\label{eq:Z2}
\end{equation}
The matrix element $\langle \tau_1 | \cM | \tau_2 \rangle$  denotes the transition amplitude 
between states corresponding to three-dimensional triangulations $\tau_1$ and $\tau_2$ in one time step. 
It is given by the sum over all four-dimensional triangulations $\cT$ of a slab with boundaries $\tau_1$ and $\tau_2$.
It depends both on the entropy factor and the action $S[\cT]$.

The probability of finding a configuration with spatial volumes $n_1, n_2, \ldots, n_T$,
\begin{equation}
  P^{(T)} (n_1, \dots, n_T) = \frac{1}{Z}  \Tr \left[| n_1 \rangle\langle n_1 | \cM | n_2 \rangle \langle n_2 | \cM | n_3 \rangle \dots  \langle n_T | \cM \right],
\label{eq:PTncM}
\end{equation}
can be measured in Monte Carlo simulations.
Here $| n \rangle\langle n | \equiv \sum_{\tau \in T_3(n)} | \tau \rangle\langle \tau | $ is a projection operator
on the subspace spanned by subset $T_3(n)$ of three-dimensional triangulations which are build of exactly $n$ three-simplices.
Although it is misleading to think of the aggregated ``state'' $| n \rangle$ as
a normalized sum of vectors $|\tau\rangle,\ \tau \in T_3(n)$,
the form of the effective action (\ref{eq:SMiniD})
suggests that there exists an effective transfer matrix $\langle n | M | m \rangle$
whose elements are labeled by the three-volumes.

For simplicity, let us define the two-point function,
\begin{equation}
	P^{(T)} (n_t, n_{t + \Delta t}) = \frac{1}{Z} \langle n_t | M^{\Delta t} | n_{t + \Delta t} \rangle \langle n_{t + \Delta t} | M^{T - \Delta t} n_t \rangle .
\label{eq:PTwoPoint}
\end{equation}
by summing over all three-volumes except at the times $t$ and $t + \Delta t$.
We will show that equation (\ref{eq:PTwoPoint}) 
provides a very good approximation to the measured data \cite{Transfer}.
The effective transfer matrix elements can be measured in various ways,
for example
$\langle n | M | m \rangle \propto \sqrt{P^{(2)} (n_1 = n, n_2 = m)}$ or $\frac{P^{(3)}(n_1 = n, n_2 = m)}{\sqrt{P^{(4)}(n_1 = n, n_3 = m)}}$.
The elements $\langle n | M | m \rangle$ 
measured in different ways completely agree up to numerical noise,
supporting equation (\ref{eq:PTwoPoint}).
All measurements presented in this paper were performed using the second expression for coupling constants $K_0 = 2.2$, $\Delta = 0.6$ and $K_4 = 0.922$.

\section{The effective action}

The minisuperspace effective action (\ref{eq:SMiniD})
is directly related to the effective transfer matrix $M$
and suggests that
\begin{equation}
\langle n | M | m \rangle = \cN e^{-\frac{1}{\Gamma} \left[ \frac{(n - m)^2}{n + m} + \mu \left( \frac{n + m}{2} \right)^{1/3} -\lambda \frac{n + m}{2} \right]} .
\label{eq:tmeff}
\end{equation}
Empirical transfer matrix elements $\langle n | M | m \rangle$
showed that for large $n_t$, up to numerical noise, approximation (\ref{eq:tmeff}) is very accurate
and allows extraction of parameters $\Gamma$, $\mu$ and $\lambda$.

\begin{figure}
\begin{center}
\includegraphics[width=0.49\textwidth]{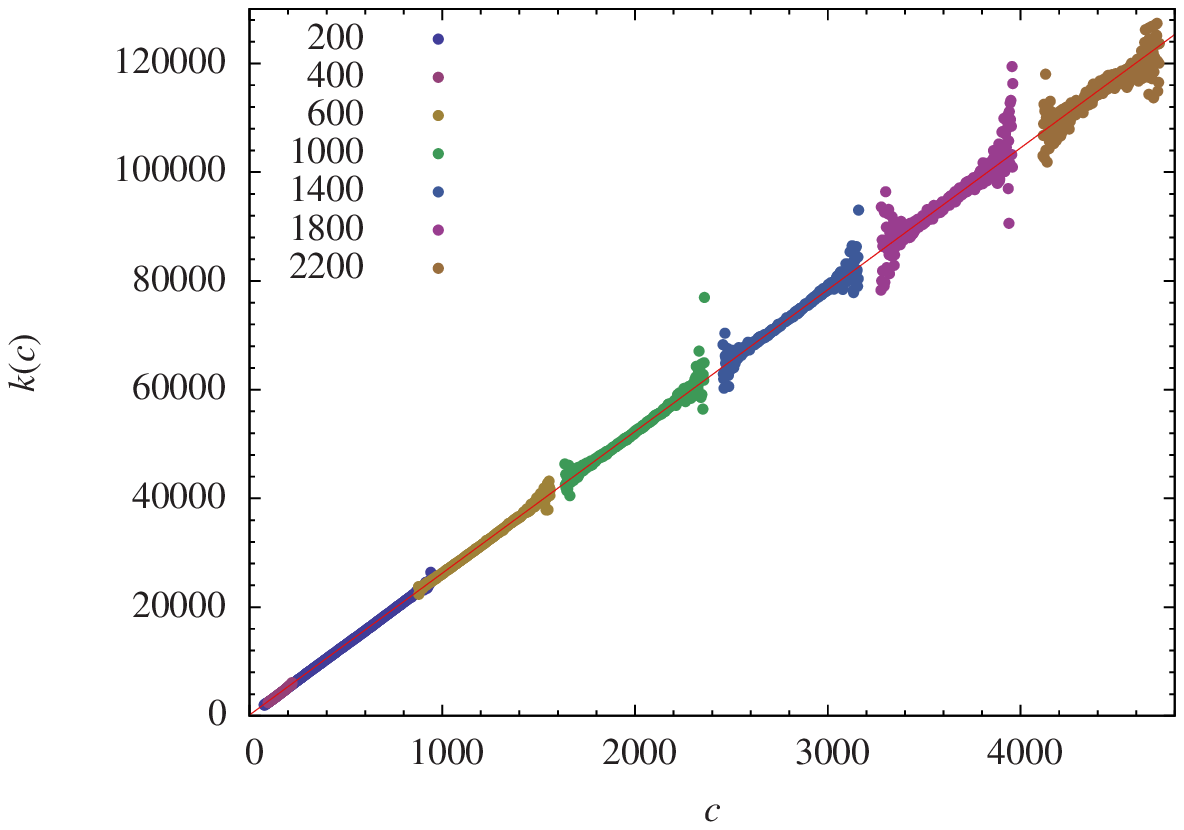}
\includegraphics[width=0.49\textwidth]{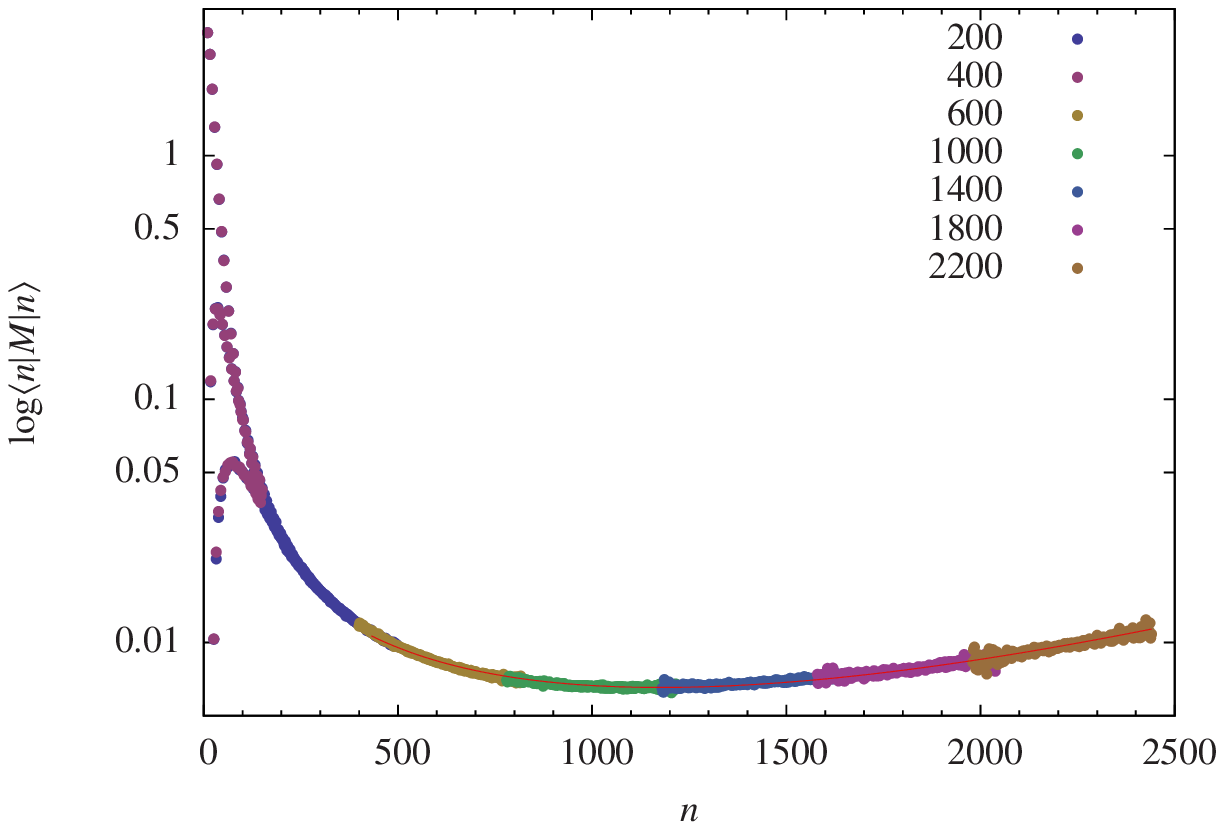}
\end{center}
\caption{
Left: The kinetic term. The coefficient $k(c)$ as a function of $c = n + m$,
and a linear fit $k(c) = \Gamma \cdot c$ (red line).
Right: The potential term.
Measured $\log \langle n | M | n \rangle$ and the fit of the potential term $-L_{eff}$ (red line,
stops at $n=400$).
}
\label{fig:kinpot}
\end{figure}

We can get a better estimation of the kinetic term
by keeping the sum of the entries $c = n + m$ fixed.
The matrix elements show the Gaussian dependence on $n$, 
$\langle n | M | c - n \rangle = 
\mathcal{N}(c) \exp \left[- \frac{(2 n - c)^2}{\Gamma \cdot c} \right] $. 
The denominator $k(c)$ of the kinetic term can be measured.
Linear behaviour $k(c) = \Gamma \cdot c$  is shown 
in Fig. \ref{fig:kinpot} (left).

The potential part of the effective Lagrangian $L_{eff}$
may be extracted from the diagonal elements of the transfer matrix,
$ L_{eff}(n, n) = - \log \langle n | M | n \rangle + \const = 
\frac{1}{\Gamma} \left( \mu n^{1/3} - \lambda n \right)$.
Right part of Fig.\ \ref{fig:kinpot} presents 
the measured effective Lagrangian and the fit.
In the bulk region, where $n_t$ is large enough, the theoretical expectation (\ref{eq:tmeff}) fits very well.
The best fit gives $\Gamma = 26.07 \pm 0.05$, $\mu = 16.5 \pm 0.2$ and $\lambda = 0.049 \pm 0.001$.
       
\section{Conclusions}

The model of Causal Dynamical Triangulations comes with a transfer matrix $\langle \tau_1 | \cM | \tau_2 \rangle$.
The measured distributions of three-volumes $n_t$, e.g. $P^{(T)}(n_t, n_{t + \Delta t})$, 
have an exact definition in terms of the full transfer matrix $\cM$ and the density matrix $| n \rangle \langle  n|$.
The data coming from Monte Carlo simulations seems to allow for a much simpler description 
in terms of an effective transfer matrix $M$,
labeled by abstract vectors $| n \rangle$ referring only to the three-volume.
The effective transfer matrix $M$ allows to directly measure the effective action $S[n_t]$.
The new method makes measurements faster.
Over the whole range of $n_t$ the effective transfer matrix elements 
are given by (\ref{eq:tmeff}) with high accuracy.
This result is fully consistent with the reduced minisuperspace action (\ref{eq:SMiniD}),
although in CDT no degrees of freedom are frozen.


\end{document}